\documentclass[10pt, 
superscriptaddress,
preprintnumbers,amsmath,amssymb,
aps,prd,nofootinbib,eqsecnum,a4paper]{revtex4}

\usepackage{graphicx}
\usepackage{epsf}
\usepackage{xcolor}

\def\be{\begin{equation}}
\def\ee{\end{equation}}
\def\bea{\begin{eqnarray}}
\def\eea{\end{eqnarray}}
\def\cs2{c_{\rm{s}}^2}

\def\PY{{\sc{Pyessence}}}

\usepackage{float}

\newcommand{\C}{\mathbb{C}}
\newcommand{\ph}{\varphi}
\newcommand\eq[1]{Eq.~(\ref{#1})}
\def\beal{\begin{align}}
\def\eeal{\end{align}}

\begin{document}

\title{Linear Density Perturbations in Multifield Coupled Quintessence}

\author{Alexander Leithes}
\address{Astronomy Unit, School of Physics and Astronomy,
Queen Mary University of London,Mile End Road, London, E1 4NS, UK}

\author{Karim A.~Malik}
\address{Astronomy Unit, School of Physics and Astronomy,
Queen Mary University of London,Mile End Road, London, E1 4NS, UK}

\author{David J. Mulryne}
\address{Astronomy Unit, School of Physics and Astronomy,
Queen Mary University of London,Mile End Road, London, E1 4NS, UK}

\author{Nelson J. Nunes}
\address{Instituto de Astrof\'{i}sica e Ci\^encias do Espa\c{c}o, Universidade de
Lisboa, 
 Faculdade de Ci\^encias, Campo Grande,
 PT1749-016 Lisboa, Portugal}

\date{\today}

\begin{abstract}
We study the behaviour of linear perturbations in multifield coupled
quintessence models. Using gauge invariant linear cosmological
perturbation theory we provide the full set of governing equations for
this class of models, and solve the system numerically.  We apply the
numerical code to generate growth functions for various examples, and
compare these both to the standard $\Lambda$CDM model and to current
and future observational bounds.
Finally, we examine the applicability of the ``small scale
approximation'', often used to calculate growth functions in
quintessence models, in light of upcoming experiments such as SKA and
Euclid. We find the deviation of the full equation results for large k modes from the approximation exceeds the experimental uncertainty for these future surveys.
The numerical code, \PY, written in Python will be publicly available.
\end{abstract}

\maketitle




\section{Introduction}
\label{Introduction}

The $\Lambda$CDM model of cosmology has become our gold standard in explaining the evolution of the universe. In this model, the dark sector of the universe is modelled by a cosmological constant, which is responsible for the acceleration of the universe in the present epoch, and a pressureless fluid that constitutes dark matter. The model is completed by assuming the presence of a baryonic matter and a radiation component. 
Remarkably, this simple picture is sufficient to explain most observational probes to date. These include high precision measurements of the CMB \cite{Adam:2015rua,ACT,SPT}, supernovae observations \cite{Perlmutter:1998np,Riess:1998cb,Kowalski:2008ez}, and large scale structure surveys \cite{Anderson:2013zyy,Bonnett:2015pww,Dawson:2015wdb}.

Despite its success, the model raises many unanswered questions such
as: Why does the cosmological constant take such an unnaturally small
value? What is the fundamental nature of dark energy? These, in
addition to other questions such as why the energy density associated
with $\Lambda$ is of the same order as that of dark matter - the
coincidence problem - have lead the community to investigate more
complex scenarios. One example is coupled quintessence. In this model
a scalar field, which makes up the dark energy (DE) component of the
universe and produces acceleration, is coupled to a pressureless dark
matter fluid
\cite{Amendola:1999dr,Holden:1999hm,Amendola:1999er,Koivisto:2005nr,Gonzalez:2006cj,Valiviita:2008iv,Amendola:2014kwa,Farrar:2003uw,Copeland:2003cv,Brookfield:2007au,Baldi:2012kt,Piloyan:2013mla,AmenTsuji,Koivisto:2015qua}.
Recent extensions which have been investigated include Multi-coupled
Dark Energy (McDE) (see e.g.~Ref.~\cite{Piloyan:2014gta}), in which
the dark matter component of the universe is formed from two fluids
that couple differently to a single scalar field.

In a series of recent papers \cite{Baldi:2012kt,Piloyan:2013mla,Piloyan:2014gta}, perturbations in the McDE model have been calculated numerically and compared with present and future large scale structure experiments. Taking this line of investigation, 
one can model the dark sector of the universe as being made up of $N$ fluids interacting with $M$ scalar fields. This model is known as Multifield or Assisted coupled quintessence \cite{Amendola:2014kwa}. The name derives from the idea that the many fields can act together to generate acceleration, in a similar manner to assisted inflation models of the early universe (see for example \cite{Liddle:1998jc,Malik:1998gy,Kanti:1999vt}). 

Multifield coupled quintessence (M$\ph$cQ) is the focus of the present paper. Our aims are two-fold. First we will calculate the equations of motion for linear perturbations in this rather general model, and incorporate these into a fast numerical code, \PY. In principal, this code can be used to generate quantities such as the growth factor of large scale structure for any coupled quintessence model with an arbitrary number of fields and fluids and arbitrary couplings. We intend to make this code publicly available. Secondly, we will apply this code, initially to revisit the McDE model, and then to consider specific models in which two scalar fields are present. Ongoing and future large scale surveys (see for example Refs.~\cite{Kitching:2015fra,Raccanelli:2015qqa}) offer a chance to distinguish between a cosmological constant and dynamical DE models, and it is important therefore to understand at what level the predictions of Multifield models will differ from those of $\Lambda$CDM and those of other quintessence models. In our work we adopt a phenomenological approach as is common in research conducted in this field. As such we have assumed that whatever the underlying particle theory may be, it includes mechanisms for screening quantum effects or other artefacts of the theory which might render it incompatible with observations. However see e.g. Refs.~\cite{D'Amico:2016kqm,Marsh:2016ynw} for possible difficulties in this approach.

For scales which are small compared to the horizon size today, an approximation to the full perturbed equations of motion has often been used in previous literature, and in particular in the previous study of McDE. A final aim of our work is to evaluate whether this approximation is sufficiently accurate, especially in the light of upcoming surveys.

The rest of this paper is set out as follows. Section \ref{IntDEBack} contains the background equations. Section \ref{GenNoGauge} contains the general gauge unspecified perturbed equations. Subsection \ref{Fixing} describes fixing the gauge in order that the equations can be solved numerically. Section \ref{PythonBack} then describes the resulting \PY~code. Section \ref{Obs} reviews the observational quantities against which our results can be compared. Finally, section \ref{PythonPert} details our numerical investigation of specific M$\ph$cQ and related models.  We conclude in Section \ref{Conclusion}.

\section{The model}

In this paper, the dark sector of the universe is modelled by $N$ different 
dark matter  fluids, with arbitrary equation of state, and $M$ different 
scalar fields. 
We also include two further fluids which model baryonic matter, and radiation. The general energy-momentum tensor for any perfect fluid is given by
\be
\label{MatSEM}
{T^{\mu}_{\nu}}^{(M_\alpha)} = (\rho_\alpha + P_\alpha) u^\mu _{(\alpha)} u_{\nu (\alpha)} + \delta^\mu_\nu P_{\alpha} \,,
\ee
where the subscript $\alpha$  labels the $N+2$ fluids, 
$\rho_\alpha$ is the density of any given fluid and 
$P_\alpha$  the corresponding pressure, and 
$u^\mu _{(\alpha)}$ is the four velocity for a given fluid.
The equation of state is defined as,
\be
\label{EOS}
w_\alpha=\frac{P_\alpha}{\rho_\alpha}\,. 
\ee
Here and throughout Greek indices $\mu$ and $\nu$ label coordinates running over time and relative dimensions in space, and we use lower case Latin indices to label only spatial dimensions. The energy-momentum tensor for the scalar fields is given by
\be
\label{FieSEM}
{T^{\mu}_{\nu}}^{(\ph)} = g^{\lambda \mu} \sum_I \partial_\lambda \ph_I \partial_\nu\ph_I - \delta^\mu_\nu \left (\frac{1}{2} \sum_I g^{\rho \sigma} \partial_\rho \ph_I \partial_\sigma \ph_I + V(\phi_1,\dots,\phi_M) \right) \,,
\ee
where $V$ is the potential energy, and upper case Roman indices label the $M$ fields. In order to model the interaction of the matter fluids with the scalar fields, we assume \cite{Amendola:1999dr,Amendola:2014kwa}

\begin{equation}
\label{econsint1}
\nabla_\mu {T^{\mu}_{\nu}}^{(\ph)} = \kappa \sum\limits_{\alpha, I} \C_{I \alpha} T_{(M_\alpha)} \nabla_\nu \ph_I \quad , \quad \nabla_\mu {T^{\mu}_{\nu}}^{(M_\alpha)} = - \kappa \sum\limits_{I} \C_{I \alpha} T_{(M_\alpha)} \nabla_\nu \ph_I \, ,
\end{equation}
where $\kappa = (8 \pi G)^{\frac{1}{2}}$ and $\C_{I \alpha}$ are coupling constants. Here $T_{(M_\alpha)}$ is the trace of Energy-momentum tensor,
\be
\label{trace}
T_{(M_\alpha)} = T^\mu_{\mu(M_\alpha)} \,, 
\ee 
for a given fluid. Equations \eq{econsint1} respect energy-momentum conservation of the total matter content. In what follows we will set the relevant components of the $C$ matrix such that there is no interaction between baryons or radiation and the scalar fields.

\subsection{Background cosmology}
\label{IntDEBack}

We take a flat Friedmann-Lema{\^\i}tre-Robertson-Walker (FLRW) spacetime as our 
background with line element 
\be
\label{BackFLRWds}
ds^2 = - dt^2 + a^2(t) \delta_{ij} dx^i dx^j \,,
\ee
where $a(t)$ is the scale factor, $t$ is cosmic time, and assume the fluids to be comoving with the expansion of the universe such that
\begin{equation}
\label{4velunpertDE}
\bar{u}_{0 (\alpha)} = -1 \qquad , \qquad \bar{u}_{i (\alpha)} = 0 \,.
\end{equation}
Here we use ``bars" to denote background quantities. The background stress energy tensor for the fluids then becomes
\bea
\label{BackSEMDE}
{\bar{T}}_{0 0} &=& \sum\limits_{\alpha} \bar{\rho}_{\alpha} + \sum\limits_{I} \frac{\dot{\bar{\ph}}_{I}^2}{2} + V \qquad , \qquad {\bar{T}}_{0 j} = 0 \qquad , \qquad {\bar{T}}_{i j} = \delta_{i j} a^2 \left(\sum\limits_{\alpha} \bar{P}_{\alpha} + \sum\limits_{I}\frac{\dot{\bar{\ph}}_{I}^2}{2} - V\right)\,,
\eea
where an overdot indicates a derivative with respect to cosmic time. \eq{econsint1} leads to the evolution equation for each fluid
\begin{equation}
\label{IntDEMatBack2}
\dot{\bar{\rho}}_{\alpha} + 3 H (\bar{\rho}_{\alpha} + \bar{P}_{\alpha}) = - \kappa \sum\limits_{I} \C_{I \alpha} (\bar{\rho}_{\alpha} - 3 \bar{P}_{\alpha}) \dot{\bar{\ph}}_{I} ,
\end{equation}
where $H$ is the Hubble parameter, and to the Klein-Gordon equation for each field
\begin{equation}
\label{IntDESFBack2}
\ddot{\bar{\ph}}_{I} + 3 H \dot{\bar{\ph}}_{I} + {V}_{, \ph_I} = \kappa \sum\limits_{\alpha} \C_{I \alpha} (\bar{\rho}_{\alpha} - 3 \bar{P}_{\alpha}) \,.
\end{equation}
The background Friedmann equation is
\be
\label{FEIntDEBack}
H^2 = \frac{\kappa^2}{3} \left[\sum\limits_{\alpha} \bar{\rho}_{\alpha} + \sum\limits_{I} \frac{\dot{\bar{\ph}}_{I}^2}{2} + V \right] \,.
\ee
Finally, we define the density parameter
\be
\label{densparam}
\Omega_\alpha = \frac{\bar{\rho}_{\alpha}}{\rho_{\rm c}} \,,
\ee
where $\rho_{\rm c}$ is the critical density defined as
\be
\label{critdens}
\rho_{\rm c} =\frac{3 H^2}{\kappa^2}\,. 
\ee

\subsection{Linear perturbations}
\label{IntDEperts}
\subsubsection{General Perturbed Equations Gauge Unspecified}
\label{GenNoGauge}
The line element for  perturbations about a  flat FLRW spacetime with the gauge unspecified is given by~\cite{Malik:2004tf}
\be 
\label{IntDEds2}
ds^2=-(1+2\phi)dt^2+2aB_{,i}dt dx^i
+a^2\left(1+2C_{ij}\right)dx^idx^j \,,
\ee
where $\phi$ is the lapse function, $B$ the shift function and partial derivatives are denoted by a ``comma". We can make the further decomposition to $C_{i j} = E_{,i j} - \psi \delta_{i j}$, where we have kept only scalar parts. The perturbed  4-velocity~\cite{Malik:2004tf} is
\be
\label{pertvelgen}
u_0 = -(1+\phi) \qquad , \qquad u_i = a(v+B),_i \,,
\ee
and the total perturbed energy-momentum tensor for our model is given by
\bea
\label{PertSEMDE}
\delta T_{0 0} &=&
  \sum\limits_{\alpha} \delta \rho_{\alpha} + \sum\limits_{I} (- \phi {\dot{\bar{\ph}}_{I}}^2 + \delta \ph_{I} \dot{\bar{\ph}}_{I} + V,_{\ph_I} \delta \ph_I) ,\\ \nonumber
 \delta T_{0 j} &=& a\left[\sum\limits_{I}\dot{\bar{\ph}}_{I}\left(\dot{\bar{\ph}}_{I} B_{,i} + \frac{1}{a} \delta \ph_{I,i}\right) - \sum\limits_{\alpha} (\bar{\rho}_{\alpha} + \bar{P}_{\alpha} )v_{(\alpha),i} \right] ,\\ \nonumber
 \delta T_{i j} &=& \delta_{i j} a^2 \left(\sum\limits_{\alpha} \delta P_{\alpha} - \sum\limits_{I}( \phi {\dot{\bar{\ph}}_{I}}^2 - \dot{\delta \ph}_{I}\dot{\bar{\ph}}_{I} + V,_{\ph_I} \delta \ph_I)  \right)\,.
\eea
Moving to Fourier space, the evolution equations for density fluctuations are given by 
\be
\label{IntDEPertEcons3}
\dot{\delta \rho}_{\alpha} - \left(\frac{k^2 v_{\alpha}}{a} +  k^2{\dot{E}} +3 \dot{\psi} \right) (\bar{\rho}_{\alpha} + \bar{P}_{\alpha} ) + 3 H (\delta \rho_{\alpha} + \delta P_{\alpha}) = - \kappa \sum\limits_{I} \C_{I \alpha} (\bar{\rho}_{\alpha} - 3 \bar{P}_{\alpha}) \dot{\delta \ph}_{I} - \kappa \sum\limits_{I} \C_{I \alpha} (\delta \rho_{\alpha} - 3 \delta P_{\alpha}) \dot{\bar{\ph}}_{I} \,,
\ee
momentum conservation gives the constraint
\be
\label{IntDEPertMomcons}
\dot{v}_{\alpha} =  \kappa \sum\limits_{I} \C_{I \alpha} (\bar{\rho}_{\alpha} - 3 \bar{P}_{\alpha}) \frac{\delta \ph_{I}}{a} + 3H \frac{\dot{\bar{P}}_{\alpha}}{\dot{\bar{\rho}}_{\alpha}} (v_{\alpha} + B) - H(v_{\alpha} + B) - \frac{\phi}{a} - \frac{\delta P_{\alpha}}{a({\bar{\rho}_{\alpha}} + {\bar{P}_{\alpha}})} - \dot{B} \,,
\ee
and the evolution of scalar field perturbations is given by
\bea
\label{IntDEPertEconsSF}
&\ddot{\delta \ph}_{I}&  + 3 H \dot{\delta \ph}_{I} + \sum\limits_{J} V,_{\ph_I \ph_J} \delta \ph_J - (k^2 {\dot{E}} + 3 \dot{\psi}) \dot{\bar{\ph}}_{I} + \frac{k^2}{a^2} \delta \ph_{I} + \frac{\dot{\bar{\ph}}_{I}}{a} k^2 B - \dot{\bar{\ph}}_{I} \dot{\phi} + 2 V,_{\ph_I} \phi\\ \nonumber &-& 2 \kappa \sum\limits_{\alpha} \C_{I \alpha} (\bar{\rho}_{\alpha} - 3 \bar{P}_{\alpha}) \phi - \kappa \sum\limits_{\alpha} \C_{I \alpha} (\delta \rho_{\alpha} - 3 \delta P_{\alpha}) = 0 \,.
\eea
The Einstein Field Equations are as follows. From the $0-0$ component we get
\be
\label{G00DEEFE}
3 H (\dot{\psi} + H \phi) + \frac{k^2}{a^2}(\psi + H[a^2\dot{E} - aB]) = - \frac{\kappa^2}{2} \left[\sum\limits_{\alpha}\delta \rho_{\alpha} + \sum\limits_{I}(- \phi \dot{\bar{\ph}}^2_{I} + \dot{\delta \ph}_{I} \dot{\bar{\ph}}_{I} + V,_{\ph_I} \delta \ph_I)\right] \,,
\ee
from the $0-i$ component 
\be
\label{Gi0DEEFE}
\dot{\psi} + H \phi = - \frac{\kappa^2}{2} \left[\sum\limits_{\alpha}a(v_{\alpha} + B)(\bar{\rho}_{\alpha} + \bar{P}_{\alpha}) - \sum\limits_{I} \dot{\bar{\ph}}_{I} \delta \ph_{I} \right] \,,
\ee
from the trace of the $i-j$ component 
\be
\label{TraceGijDEEFE}
\ddot{\psi} + 3 H \dot{\psi} + H \dot{\phi} + (3 H^2 + 2 \dot{H}) \phi = \frac{\kappa^2}{2} \left[\sum\limits_{\alpha}\delta P_{\alpha} + \sum\limits_{I} (- \phi \dot{\bar{\ph}}^2_{I} + \dot{\delta \ph}_{I} \dot{\bar{\ph}}_{I} - V,_{\ph_I} \delta \ph_I)\right] \,,
\ee
and from the trace-free part of the $i-j$ component 
\be
\label{TraceFreeGijDEEFE}
{\dot{\sigma}}_s + H \sigma_s - \phi + \psi = 0 \,,
\ee
where $\sigma_s$ is the scalar shear and $\sigma_s = a^2 \dot{E} - a B$.

\subsubsection{Governing equations in flat gauge}
\label{Fixing}

Many gauge choices are available. Previously in the literature a common choice of gauge for studies of coupled quintessence models has been the longitudinal gauge ($\tilde{B} = \tilde{E} = 0$), and we present the equations of motion for perturbations in this gauge in Appendix~\ref{Long2f2dm}. However, we found that this gauge is not a good choice for the numerical integration of the full equations of motion.  This is due to the prefactor term in \eq{Phiconstraint}. The magnitude of the second term in this prefactor is orders of magnitude smaller than the first, except when the first touches zero, which can occur as the fields oscillate. This leads to a loss of accuracy at these times and to a numerical instability.  For our numerical integration we therefore use the flat gauge which does not suffer from this problem.

The flat gauge is defined by the conditions $\tilde{\psi}=0$ and $\tilde{E}=0$. Defining the new quantity
\be
\label{FlatNewv}
\hat{v}_\alpha = v_\alpha + B\,,
\ee
in this gauge, 
\eq{IntDEPertEcons3} reduces to
\be
\label{FlatIntDEPertEcons4}
\dot{\delta \rho_\alpha} + 3 H (\delta \rho_\alpha + \delta P_\alpha) - \frac{k^2 (\hat{v}_\alpha - B)}{a}({\bar{\rho}}_\alpha + {\bar{P}}_\alpha) = - \sum\limits_{I} \kappa \C_{I \alpha} ({\bar{\rho}}_\alpha - 3 {\bar{P}}_\alpha) {\dot{\delta \ph}}_I - \sum\limits_{I} \kappa \C_{I \alpha} (\delta \rho_\alpha - 3 \delta P_\alpha) {\dot{\bar{\ph}}}_I \,,
\ee
and \eq{IntDEPertMomcons} to
\be
\label{FlatIntDEPertMomconsLong}
\dot{\hat{v}}_\alpha =  \kappa \sum\limits_{I} \C_{I \alpha} (\bar{\rho}_\alpha - 3 \bar{P}_\alpha) \frac{\delta \ph_I}{a} + 3H \frac{\dot{\bar{P}}_\alpha}{\dot{\bar{\rho}}_\alpha} \hat{v}_\alpha - H\hat{v}_\alpha - \frac{\phi}{a} - \frac{\delta P_\alpha}{a({\bar{\rho}_\alpha} + \bar{P}_\alpha)}\,.
\ee
The evolution equation for the fields, \eq{IntDEPertEconsSF}, becomes
\bea
\label{FlatDEPertEconsSF2}
&{\ddot{\delta \ph}}_I& + 3 H {\dot{\delta \ph}}_I + \sum\limits_{J} V,_{\ph_I \ph_J} \delta \ph_J - \left[ \frac{\kappa^2}{2H} \left( \sum\limits_{\alpha} \delta P_\alpha - \sum\limits_{I} (\phi \dot{\bar{\ph}}^2_I - {\dot{\delta \ph}}_I {\dot{\bar{\ph}}}_I + V,_{\ph_I} \delta \ph_I ) \right) - \frac{(3 H^2 + 2 \dot{H})}{H} \phi \right] {\dot{\bar{\ph}}}_I \\ \nonumber &+& \frac{k^2}{a^2} \delta \ph_I + \frac{k^2 B}{a} {\dot{\bar{\ph}}}_I + 2 V,_{\ph_I} \phi - 2 \sum\limits_{\alpha} \kappa \C_{I \alpha} ({\bar{\rho}}_\alpha - 3 {\bar{P}}_\alpha) \phi - \sum\limits_{\alpha} \kappa \C_{I \alpha} (\delta \rho_\alpha - 3 \delta P_\alpha) = 0 \,. 
\eea
From \eq{G00DEEFE}, we get
\be
\label{FlatG00DEEFE2}
3 H^2 {\phi} - \frac{k^2 B}{a} H = - \frac{\kappa^2}{2} \left[\sum\limits_{\alpha} \delta \rho_\alpha + \sum\limits_{I}(-\phi \dot{\bar{\ph}}^2_I + {\dot{\delta \ph}}_I {\dot{\bar{\ph}}}_I + V,_{\ph_I} \delta \ph_I)\right] \,,
\ee
and from \eq{Gi0DEEFE} 
\be
\label{FlatGi0DEEFE2}
\phi = - \frac{\kappa^2}{2H} \left[\sum\limits_{\alpha} a \hat{v}_\alpha (\bar{\rho}_\alpha + \bar{P}_\alpha) - \sum\limits_{I} {\dot{\bar{\ph}}}_I \delta \ph_I \right] \,,
\ee
which allows us to replace  $\phi$ in terms of field and fluid perturbations. For completeness we note that \eq{TraceGijDEEFE} gives
\be
\label{FlatTraceGijDEEFE2}
H \dot{\phi} + (3 H^2 + 2 \dot{H}) \phi = \frac{\kappa^2}{2} \left[\sum\limits_{\alpha} \delta P_\alpha - \sum\limits_{I} \left(\phi \dot{\bar{\ph}}^2_I - {\dot{\delta \ph}}_I {\dot{\bar{\ph}}}_I + V,_{\ph_I} \delta \ph_I\right) \right] 
\ee
and from \eq{TraceFreeGijDEEFE} we have
\be
\label{FlatTraceFreeGijDEEFE2}
\dot{B} + 2HB  = - \frac{\phi}{a} \,.
\ee
Combining \eq{FlatG00DEEFE2} and \eq{FlatGi0DEEFE2} we find 
\be
\label{FlattildeBNewVar}
B = \frac{3 \kappa^2 a}{2k^2} \left[ \frac{1}{3H} \left( \sum\limits_{\alpha} \delta \rho_\alpha - \sum\limits_{I} (\phi \dot{\bar{\ph}}^2_I - {\dot{\delta \ph}}_I {\dot{\bar{\ph}}}_I - V,_{\ph_I} \delta \ph_I ) \right) + \sum\limits_{I} {\dot{\bar{\ph}}}_I \delta \ph_I - \sum\limits_{\alpha} a \hat{v}_\alpha (\bar{\rho}_\alpha + \bar{P}_\alpha) \right] \,,
\ee
which allows us to replace $B$ is terms of field and fluid perturbations.

\section{Numerical solutions}
\label{PythonBack}

We can now solve the closed system of equations derived in the previous section numerically. The system of background equations for the scalar fields and the energy densities of the fluids, \eq{IntDEMatBack2} and \eq{IntDESFBack2}, together with the Friedmann constraint \eq{FEIntDEBack}, is solved simultaneously with the evolution equations for the perturbations $\delta\rho_\alpha$, $\hat v_\alpha$ and $\delta \varphi_I$, \eq{FlatIntDEPertEcons4} to \eq{FlatDEPertEconsSF2}, together with the constraint equations for $\phi$ and $B$, \eq{FlatGi0DEEFE2} and \eq{FlattildeBNewVar}. The numerical code, named \PY, is written in Python and publicly available on Bitbucket~\cite{Bitbucket} and on the \PY~website~\cite{Pyweb} under an open source modified BSD license, with documentation available in Ref.~\cite{PYDOCREF}.

\subsection{Initial Conditions}
\label{CQIC}

\subsubsection{Background Initial Conditions}

We set the initial conditions for the background energy densities of the fluids and the background field amplitudes such that the background evolution follows closely that of the $\Lambda$CDM model. Due to the potentials used in the models tested we have analytical solutions for the background evolution equations, which enables us to set the background initial conditions in terms of their values today. We are free to choose an initial time, and select $N=-14$, which fixes the initial value for the scale factor $a$ and coordinate time, $t$. This also ensures we are well into the radiation dominated epoch. In particular, we demand that the model satisfies constraints on present day energy densities from Planck data \cite{Ade:2015xua}. These are $\Omega_{\Lambda} = 0.6911 \pm 0.0062$ for the cosmological constant, $\Omega_{r} = 9.117 \times 10^{-5}$ for radiation, $\Omega_{b} = 0.0486 \pm 0.0003$ for baryons and $\Omega_{CDM} = 1 - \Omega_{DE} - \Omega_{r} - \Omega_{b}$ for cold dark matter. To do so, we assume that the scalar fields will collectively replace $\Lambda$, and the dark matter fluids collectively replace the single cold dark matter species of the $\Lambda$CDM model. Initially we take the fields' velocity to be zero, $\dot \varphi_I=0$. Of course we need to check on a case by case basis whether the fields really do generate acceleration in a way that accounts for observations, and that dark matter components behave in a viable way, such that the background evolution is compatible with current limits.

\subsubsection{Perturbed Initial Conditions}
\label{perticflat}

We start our simulations at sufficiently early times to ensure radiation domination and that all the $k$ modes studied lie outside the horizon at that time. For simplicity, we choose the initial conditions for the field velocity and  field perturbations to be zero
\be
{\dot{\delta \ph}}_I =  \delta \ph_I = 0 \,,
\ee
though we find the evolution is insensitive to this choice. The initial conditions for all other perturbations can be given in terms of observational constraints on the power spectrum of the gauge invariant curvature perturbation $\zeta$, (see for example Ref.~\cite{Liddle:2000cg}), 

\be
\label{zetadef}
\left\langle\zeta^2\right\rangle = \delta^3 ({\bf{k-k}}') \frac{2 \pi^2}{k^3} {\cal{P}}_\zeta (k) \,,
\ee
where $\zeta$, the curvature perturbation on uniform density hypersurfaces, is defined as
\be
\label{zetadefn}
-\zeta  = \psi + \frac{H}{\dot{\bar{\rho}}} \delta \rho \,.
\ee
On superhorizon scales the power spectrum can be parametrised as 
\be
\label{Ps}
{\cal{P}}_\zeta (k) = A_s \left( \frac{k}{k_*} \right)^{n_s - 1} \,,
\ee
where \cite{Ade:2015lrj} $ A_s = 2.142 \times 10^{-9}$ is the scalar amplitude at the Planck pivot scale
$k_* = 0.05$ Mpc$^{-1}$,
and $n_s=0.9667$ is the spectral index~\cite{Ade:2015xua}. \\
From \eq{zetadefn} we then get a relation between the curvature perturbation and the total energy density perturbation in flat gauge, such that,
\be
\delta\rho_{\rm flat}=-\frac{\dot{\bar\rho}}{H}\zeta\,.
\ee
This allows us to set the initial condition for the individual fluids. In addition we assume that the initial conditions are adiabatic, which gives a relation between the fluid density perturbations initially. The gauge-invariant relative entropy perturbation between two non-interacting fluids~\cite{Malik:2004tf} is given by
\be
\label{entpert}
{\cal{S}}_{\alpha \beta} = 
- 3 H \left( \frac{\delta \rho_\alpha}{\dot{\bar{\rho}}_\alpha} - \frac{\delta \rho_\beta}{\dot{\bar{\rho}}_\beta} \right) \,.
\ee
Adiabatic initial conditions require that ${\cal{S}}_{\alpha \beta} =
0$. Combining \eq{entpert} with \eq{IntDEMatBack2} for radiation and
baryons, which for these models, as specified in
Section~\ref{Introduction} have couplings of zero, we find \be
\label{deltarels}
\delta_{b} = \frac{3}{4} \delta_r \,,
\ee
where we introduced the density contrast for a given fluid species, $\alpha$, as
\be
\label{denscontdefn}
\delta_\alpha 
\equiv \frac{\delta \rho_\alpha}{\bar{\rho}_\alpha}  \,.
\ee
Finally we can set the initial conditions for the 3-velocities, $\hat{v}_\alpha$.  We checked numerically that the late time evolution of the system is not very sensitive to the actual value for the 3-velocities, and we therefore set $\hat{v}_\alpha=0$ initially. While studying the initial conditions we found that aside from the initial radiation density perturbation the results are fairly insensitive to small changes in the initial conditions, due to the integration starting well inside radiation domination. Small variations in the initial conditions for the other constituents, for a given $k$ mode, soon converged to a common trajectory within approximately one e-fold from the start of the simulations. This meant there was negligible difference in the observable growth of the density perturbations.

\subsubsection{Relating Longitudinal Gauge to Flat Gauge}
\label{pertininlong}

In the previous sections we have presented the system of governing equations and the initial conditions for the code in flat gauge. However, in order to connect to previous studies in the literature we present our results in terms of the density contrast in longitudinal gauge.\\
Using the background and perturbed densities as defined in \eq{BackSEMDE} and \eq{PertSEMDE}, the total density contrast is defined as,
\be
\label{totdenscontdefn}
\delta = \frac{\sum\limits_{\alpha} \delta \rho_\alpha + \sum\limits_{I} \delta \rho_{\ph_I}}{\sum\limits_{\alpha} \bar{\rho}_\alpha + \sum\limits_{I} \bar{\rho}_{\ph_I}}  \,.
\ee
Using the transformations for the metric and matter variables given in appendix~\ref{Flat to Long Gauge Relations}, and the constraint Eqns.~(\ref{FlattildeBNewVar}), we find
\be
\label{KAMdeltalinkbulk}
\delta_{\rm long}=\delta_{\rm flat} + \frac{{\dot{\bar{\rho}}}^2}{\bar\rho} a \left( \frac{3 \kappa^2 a}{2k^2} \left[ \frac{1}{3H} \left( \sum\limits_{\alpha} \delta \rho_\alpha - \sum\limits_{I} (\phi \dot{\bar{\ph}}^2_I - {\dot{\delta \ph}}_I {\dot{\bar{\ph}}}_I - V,_{\ph_I} \delta \ph_I ) \right) + \sum\limits_{I} {\dot{\bar{\ph}}}_I \delta \ph_I - \sum\limits_{\alpha} a \hat{v}_\alpha (\bar{\rho}_\alpha + \bar{P}_\alpha) \right] \right)  \,,
\ee
which reduces initially to
\be
\label{KAMdeltalinkIC}
\delta_{\rm long}=\delta_{\rm flat}
+\left(\frac{k}{a}\right)^{-2}\left[4\pi\, G\delta_{\rm flat}
-\frac{{\dot{\bar{\rho}}}^2}{3H\bar\rho}
a\sum_\alpha (\bar{\rho}_\alpha + \bar{P}_\alpha)   \hat v_\alpha\right]   
\,.
\ee

\section{Observations}
\label{Obs}

Two key parameters which are constrained by observational data are the growth factor and growth function. We therefore apply our code to calculate these quantities. The growth factor is defined as 
\be
\label{growthdef}
g = \frac{\delta}{\delta_0} \,,
\ee
where $\delta$ is the total density contrast defined in the longitudinal gauge \cite{Raccanelli:2015qqa}, 
and $\delta_0$ is the total density contrast today. The growth function, $f$, is defined as
\be
\label{fdefn}
f = \frac{\delta'}{\delta} \,,
\ee
where the prime in this case denotes a derivative with respect to the number of e-folds \cite{Raccanelli:2015qqa}. 
Typically observational results are presented as constraints on the combinations $fg$ and $f\sigma_8$, since, for example, these quantities can be extracted directly from redshift space distortions  (see e.g.~\cite{Bull:2015lja}). $\sigma_8$ is the amplitude of the matter power spectrum at a scale of $8h^{-1}$Mpc \cite{Raccanelli:2015qqa,Macaulay:2013swa}. The experimental uncertainty of $\sigma_8$, taken from DES, which overlaps two other data sets which are in some tension (CFHTLenS and Planck), is $0.81^{+0.16}_{-0.26}$\cite{Abbott:2015swa}. In Subsection~\ref{Lspaceexplore} we use $\sigma_8 = 0.81$~\cite{Ade:2015xua} since this is consistent with the other Planck based parameter values we have used. Future surveys hope to have the sensitivity to pick up $k$ dependence in the growth of structure. SKA~\cite{Bull:2015lja,Raccanelli:2015qqa}, for example, should be sensitive to measurements of growth at approximately the percent level (or better) for $42H_0 < k < 420H_0$ at a redshift $z \approx 1$~\cite{Bull:2015lja}. For $k > 42H_0$ this sensitivity falls to $\approx 30\%$, for example, being at this level around $k = 21H_0$. According to the author~\cite{Bull:2015lja} this combined four survey approach (SKA1-MID Band 1 and Band 2 IM (intensity mapping) surveys, H$\alpha$ and SKA2) should therefore have sufficient accuracy to distinguish between GR (General Relativity) + $\Lambda$CDM and alternative models, such as coupled quintessence. This accuracy is potentially increased still further through multiple tracer analysis, cross-correlating with other surveys such as Euclid. The combined redshift range for SKA and Euclid is $0.5 \lesssim z \lesssim 2$.

Current surveys offer far looser constraints on the growth of structure. Below we use observational data from 6dFGS, LRG$_{200}$, LRG$_{60}$, BOSS, WiggleZ and VIPERS with associated errors~\cite{Macaulay:2013swa} in our plots for $fg$. These current surveys have a shorter redshift range ($z \lesssim 0.8$) and constrain growth at only $\approx 10 - 20 \%$ level. In single field coupled quintessence there is an observational constraint on the magnitude of the coupling between DE and CDM as $ \C < 0.1 \sqrt{\frac{2}{3}}$~\cite{Amendola:1999er}. For this class of models couplings greater than this give unrealistic background cosmologies, through deviations in the sound horizon at decoupling from that obtained in $\Lambda$CDM (see e.g.~\cite{Amendola:1999er}). The McDE models first described in Section~\ref{McDE} (1 scalar field and 2 CDM species) give viable background cosmologies through the effect of the opposite charges and symmetric magnitudes of the CDM species~\cite{Piloyan:2014gta}. We restrict our background analysis to ensure that the relative background densities match today's values, and that the evolution moves from radiation domination, through a period of CDM domination to a final epoch of DE domination.

\section{Example models}
\label{PythonPert}

\begin{figure}
\centerline{\includegraphics[angle=0,height=72mm]{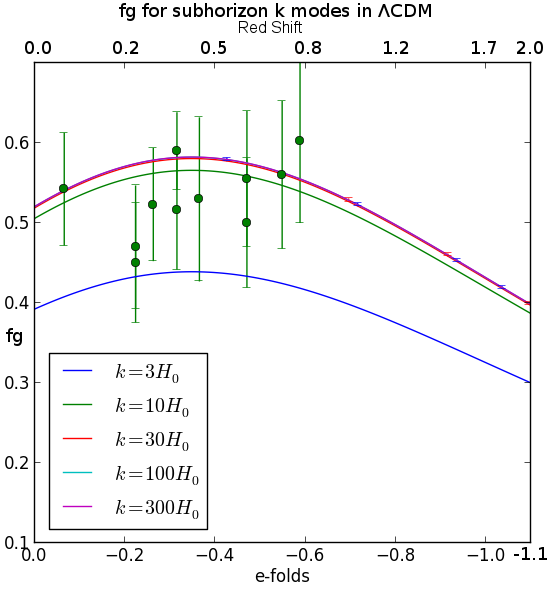}\includegraphics[angle=0,height=72mm]{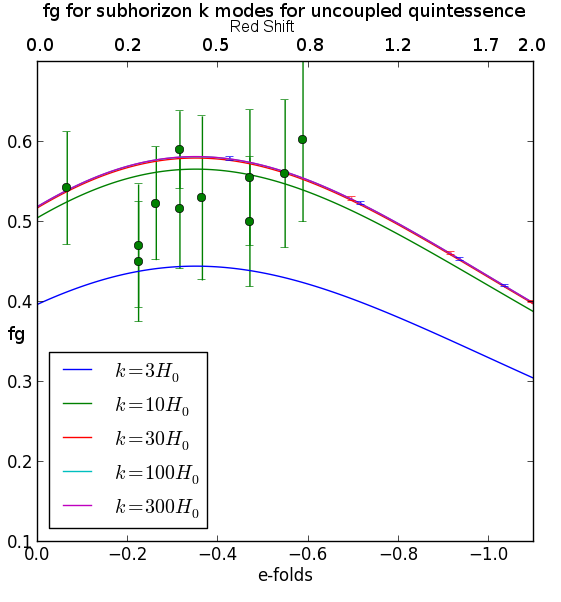}\includegraphics[angle=0,height=72mm]{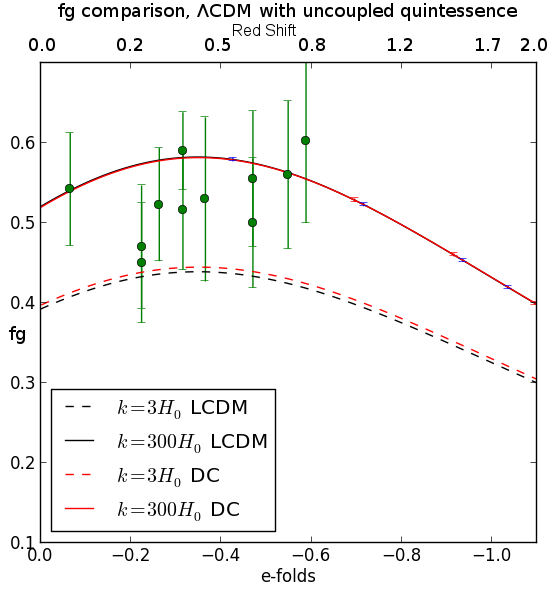}}
\caption{The left plot shows the growth function, $fg$, on sub-horizon scales for $\Lambda$CDM, for the region of redshifts relevant for current and future surveys. The green points are observational data from 6dFG
S, LRG$_{200}$, LRG$_{60}$, BOSS, WiggleZ and VIPERS with associated errors~\cite{Macaulay:2013swa}. The red error bars are the Euclid forecasts and the blue the SKA forecasts~\cite{Raccanelli:2015qqa} applied to the $k=300H_0$ plot. The forecast error bars are approximately the line width. The centre plot shows the same for uncoupled two field two CDM species quintessence, $\lambda=0.1$. The right hand plot compares $fg$ for $\Lambda$CDM with uncoupled quintessence (DC) for $k=300H_0$ and $k=3H_0$.}
\label{fLCDMsubhorLONG}
\end{figure}

In order to compare models against the standard model, we first applied our code to produce results for the $\Lambda$CDM cosmology. Figure~\ref{fLCDMsubhorLONG} shows the results for the behaviour of $fg$ together with current observational constraints. We also applied our code to a uncoupled quintessence model with two scalar fields and two CDM species. In this case, and for all subsequent models including McDE, the potential for the scalar fields is taken to be a sum of exponentials,
\be
\label{sumofexppot}
V(\ph_1,\ldots,\ph_I) = M^4 \sum\limits_{I} e^{-\kappa \lambda_I \ph_I} \,,
\ee
where $\lambda_I$ is the slope of the potential for field $I$ and $M$ is the scale of the potential. These results are also shown in Figure~\ref{fLCDMsubhorLONG}. We can see that for large $k$ there is a negligible difference in the growth, and even for small $k$, the difference is still too small to be detectable by future surveys such as SKA and Euclid.

\subsection{Multi-coupled Dark Energy - McDE}
\label{McDE}
Next, we investigated the recently proposed subclass of coupled quintessence, McDE, as described in Refs.~\cite{Baldi:2012kt,Piloyan:2013mla,Piloyan:2014gta}. The McDE model has two CDM species coupled to one DE scalar field. The couplings of each DM species have the same magnitude but opposite signs. In order to compare directly with the results of Ref.~\cite{Piloyan:2014gta}, we set the baryon density to zero for this model. In previous work, perturbations in this model have been studied using an approximation to the full system of equations \cite{Piloyan:2014gta,Amendola:2014kwa,AmenTsuji,Amendola:1999er}. This simplification is valid for modes on subhorizon scales and allows scalar field fluctuations to be written in terms of density perturbations. The dimensionality of the system can therefore be reduced and an autonomous system of equations formed for the density perturbations alone. We use the system of ODEs, taken from Ref.~\cite{Piloyan:2014gta}, to evolve the density perturbations. We also use the same initial conditions to generate results using our implementation of the full equations. This provides a useful examination of the applicability of the subhorizon approximation. Finally, for comparison, we produce $\Lambda$CDM results with the assumption of zero baryonic content, using the McDE subhorizon approximations equations and our full system of equations.

We take the initial conditions used in Figure 7 of Ref.~\cite{Piloyan:2014gta}. The couplings are symmetric and set to $\beta = \pm 0.03$ where $\beta \equiv \left(\sqrt{\frac{3}{2}}\right) \C$ and $\alpha=0.12$ where $\alpha \equiv \lambda$. The potential is as \eq{sumofexppot}, for $I=1$, $\alpha=2$. The initial conditions were set non-adiabatically with $A_{IC}=2$, where  
\be
\label{adiab}
A_{IC} = \frac{\Omega_- \delta_{-i}}{\Omega_+ \delta_{+i}}\,, 
\ee
and $A_{IC}$ is the measure of the deviation from adiabaticity, `$-$' denote the negatively charged CDM species and `$+$' the positively charged. One further parameter is the asymmetry between these two species, $\mu$, and is defined
\be
\label{mu}
\mu=\frac{\Omega_+ - \Omega_-}{\Omega_+ + \Omega_-}\,. 
\ee
Initially $\mu=0.5$, however we found the final results to be insensitive to this initial condition. Once again we generated plots using the reduced system and the full equations for a range of $k$s. For quantities which were absent in \cite{Piloyan:2014gta}; radiation perturbations, perturbations to the scalar field, these were initially set to zero.

\begin{figure}
\centerline{\includegraphics[angle=0,height=70mm]{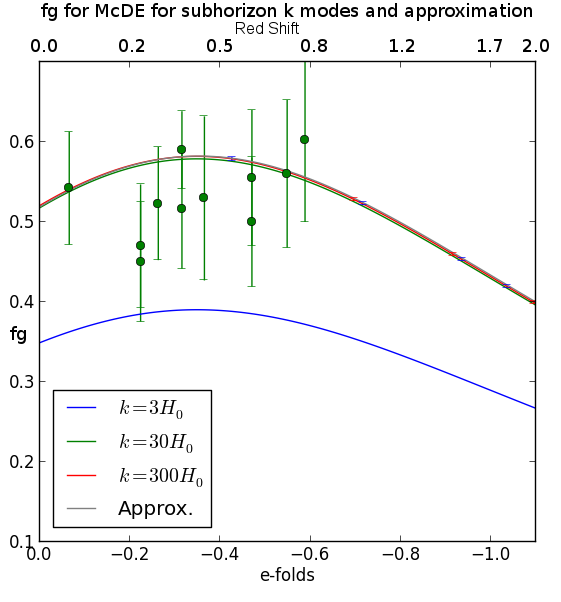}\includegraphics[angle=0,height=70mm]{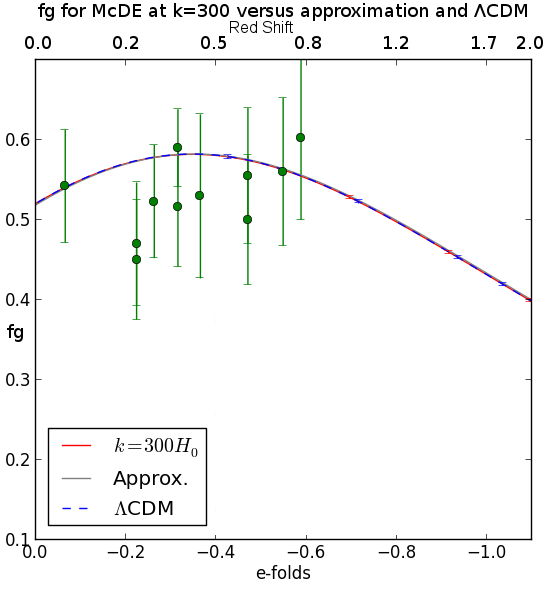}\includegraphics[angle=0,height=70mm]{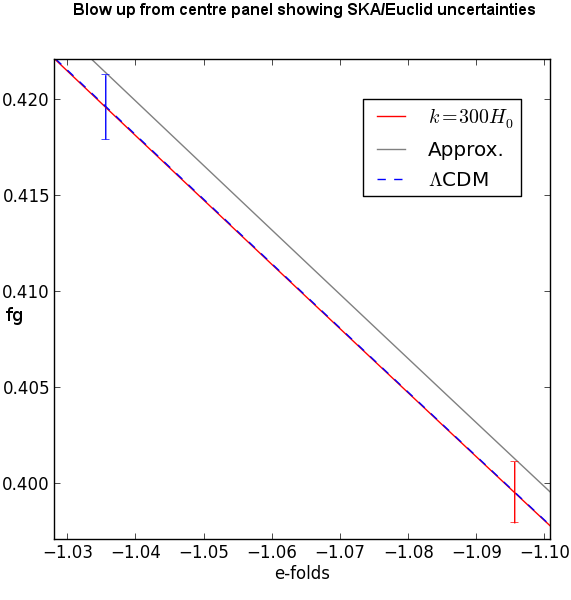}}
\caption{The left hand panel shows $fg=\frac{\delta'}{\delta_0}$ for McDE with $\Omega_{\Lambda}=0.692$, no baryons, one CDM species and unperturbed radiation, $\lambda=0.12$, $\C=\pm0.03\sqrt{\frac{2}{3}}$. A range of subhorizon $k$ modes are shown with convergence towards a $k$ independent evolution of growth with larger $k$s. The result for the subhorizon approximation from Ref.~\cite{Piloyan:2014gta} is shown in grey. The centre panel shows $fg$ for McDE for $k=300H_0$ for the full equations,  the subhorizon approximation from Ref.~\cite{Piloyan:2014gta} and $\Lambda$CDM for $k=300H_0$. In each panel, the green points are observational data from 6dFGS, LRG$_{200}$, LRG$
_{60}$, BOSS, WiggleZ and VIPERS with associated errors~\cite{Macaulay:2013swa}. The red error bars are the Euclid forecasts and the blue the SKA forecasts~\cite{Raccanelli:2015qqa} applied to the $k=300H_0$ plot. The right panel reproduces a magnified area of the centre panel, showing that the approximation results differ from the full equations by more than the uncertainties.}
\label{fgLCDMBALDI}
\end{figure}

The results are presented in terms of the evolution of $fg$ and are shown in Figure~\ref{fgLCDMBALDI}. For the simplified $\Lambda$CDM model, with the baryon content set  to zero,  and the radiation unperturbed (initially for our full code, while radiation perturbation equations are not included in the subhorizon approximation) the results are shown in Figure~\ref{fgLCDMBALDI} together with present and future constraints.
Examining this figure, we see that for the largest $k$ modes the results converge with the result generated using the subhorizon approximation. It should be noted however that there is a noticeable difference in the evolution of growth between the different $k$ modes down to the scale of $k=300H_0$, and as such the subhorizon approximation is masking this $k$ dependence over this range of $k$s.

As in Ref.~\cite{Piloyan:2014gta} we found that the evolution provided by the subhorizon approximation gives an evolution for $fg$ close to $\Lambda$CDM but with a deficit at lower red shifts. The larger $k$ modes have mostly converged with the approximation, however, there is a small deviation such that at late times $fg$ is closer to $\Lambda$CDM than the approximations. As with all full equation results produced, the growth results are converging with increasing $k$, as expected. However, even at scales of $k=300H_0$ the small scale approximation appears insufficient for this model, even for the conservative predicted precision for SKA and Euclid measurements. We can see in the right hand plot of Figure~\ref{fgLCDMBALDI} that the approximation deviates from the full equations results by more than the predicted observational precision at these higher redshifts. Additionally, for the full equations at $k=300H_0$ the evolution of $fg$ for McDE and $\Lambda$CDM models can not be distinguished from the predicted observational precision.

\subsection{Multifield Coupled Quintessence}
\label{ACQ}
\subsubsection{Transient Matter Domination}
\label{CQSS}

\begin{figure}
\centerline{\includegraphics[angle=0,width=80mm]{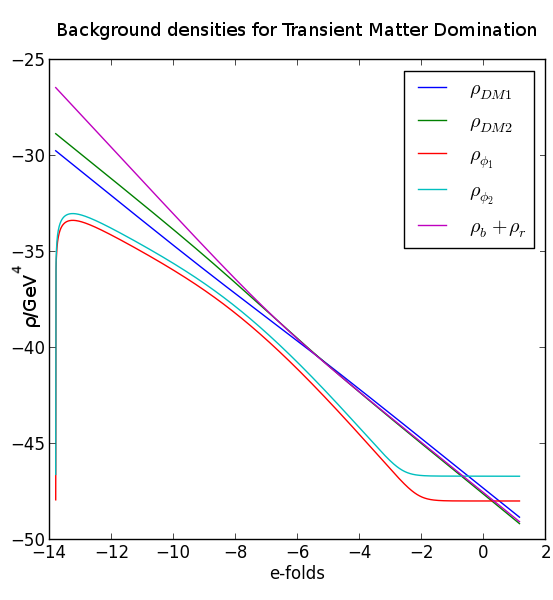}\includegraphics[angle=0,width=80mm]{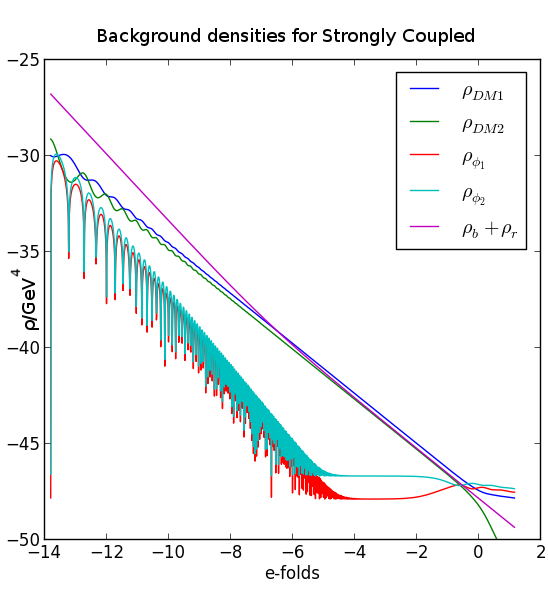}} 
\caption{The left hand plot shows the evolution of the background densities of components for the transient matter domination M$\ph$cQ model. The scale is logarithmic. Subscript $b$ denotes baryons, subscript $r$ denotes radiation. Couplings, $\C_{11}=-0.2$, $\C_{12}=0.4$, $\C_{21}=-0.3$, $\C_{22}=0.6$. Slopes for the potentials, $\lambda_1 = \lambda_2 = 0.1$. The right hand plot shows the evolution of the background densities of components for the strongly coupled matter dominated coupled quintessence model. Subscript $b$ denotes baryons, subscript $r$ denotes radiation. Couplings, $\C_{11}=-20$, $\C_{12}=40$, $\C_{21}=-30$ and $\C_{22}=60$. Slopes for the potentials, $\lambda_1 = \lambda_2 = 10$}
\label{backmatdom}
\end{figure}

Next we considered the M$\ph$cQ model introduced in Ref.~\cite{Amendola:2014kwa}. The model contains two pressureless dark matter fluids coupled to two scalar fields. Initially we choose small couplings ($\C_{11}=-0.2$, $\C_{12}=0.4$, $\C_{21}=-0.3$ and $\C_{22}=0.6$) and small slopes for the potentials, $\lambda_I$, ($\lambda_1 = \lambda_2 = 0.1$). The evolution of the background densities for this model is shown in the left hand panel of Figure~\ref{backmatdom}. These small couplings give rise to a tracking behaviour, by which we mean that the scalar fields densities between $e$-folds of within the interval  $-13$ and $-3$ approximately follow the evolution of the energy densities of the other components. This may alleviate the coincidence problem.
This model also still gave a transition to a near constant energy density for the scalar fields at late times and domination of the scalar field energy densities at late times, as required to produce similar background behaviour to $\Lambda$CDM.

\begin{figure}
\centerline{\includegraphics[angle=0,height=70mm]{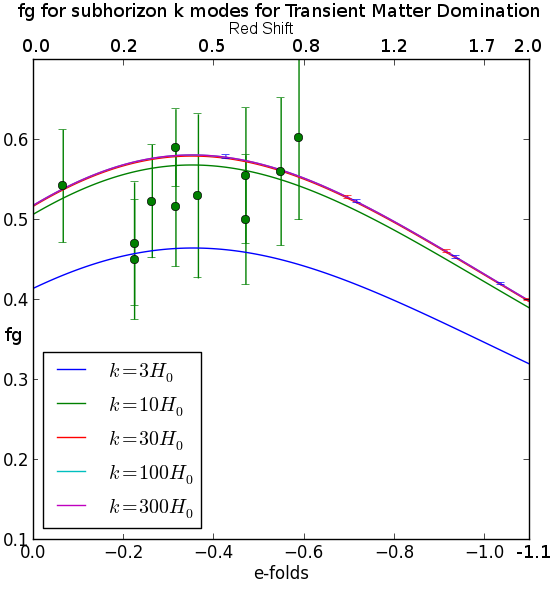}\includegraphics[angle=0,height=70mm]{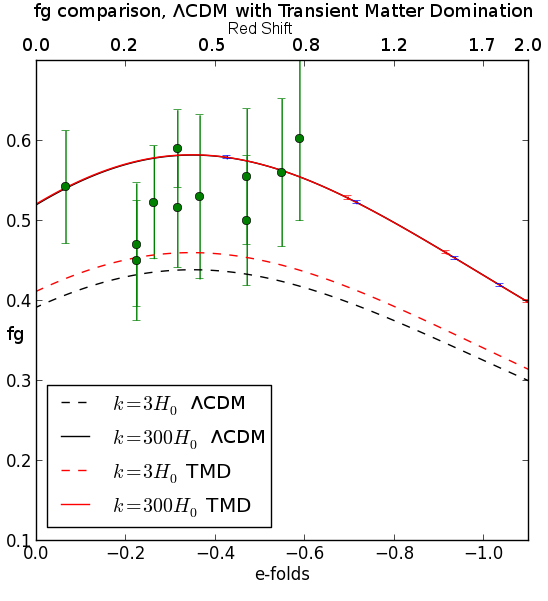}\includegraphics[angle=0,height=70mm]{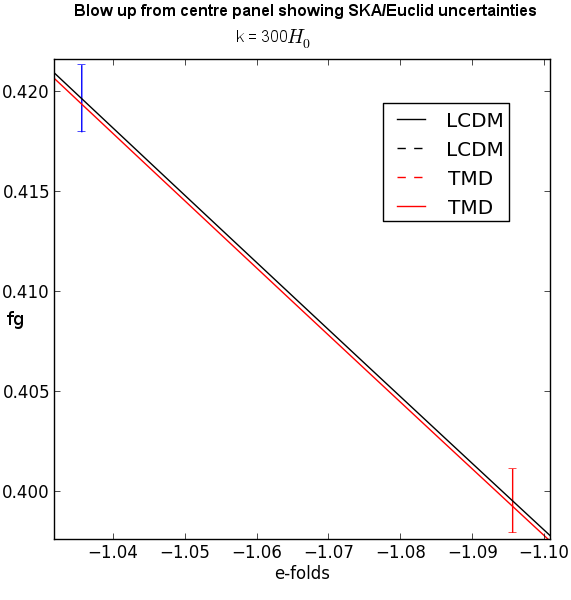}} 
\caption{The left plot shows the growth function, $fg$, sub-horizon scales, for the transient matter domination M$\ph$cQ model, for the region of redshifts relevant for current and predicted future surveys. Couplings, $\C_{11}=-0.2$, $\C_{12}=0.4$, $\C_{21}=-0.3$, $\C_{22}=0.6$. Slopes for the potentials, $\lambda_1 = \lambda_2 = 0.1$.  The green points are observational data from 6dFGS, LRG$_{200}$, LRG$_{60}$, BOSS, WiggleZ and VIPERS with associated errors~\cite{Macaulay:2013swa}. The red error bars are the Euclid forecasts and the blue the SKA forecasts~\cite{Raccanelli:2015qqa} applied to the $k=300H_0$ plot. The centre plot compares the $fg$ between $\Lambda$CDM and transient matter dominated model (TMD) for $k=300H_0$ and $k=3H_0$. The right hand panel zooms in on the centre panel to show the results versus the SKA/Euclid uncertainties for $k=300H_0$.}
\label{fCQMDsubhorLONG}
\end{figure}

The right hand panel of Figure~\ref{fCQMDsubhorLONG} is the evolution of $fg$ for $k=300H_0$, and shows the conservative predicted observational precision would not be enough to distinguish between this model and $\Lambda$CDM. However, if optimal performance were achieved leading to an order of magnitude improvement in the observational uncertainties this could be sufficient to distinguish the two models.

\subsubsection{Strongly Coupled Matter Domination}
\label{CQNO}
Taking again the same setup, next we choose the couplings  $\C_{11}=-20$, $\C_{12}=40$, $\C_{21}=-30$ and $\C_{22}=60$ and the slopes for the potentials  $\lambda_1 = \lambda_2 = 10$. The background evolution of this system was also studied in Ref.~\cite{Amendola:2014kwa} and can be seen in the right hand plot in Figure~\ref{backmatdom}. The initial oscillations in the scalar fields are caused by the initial conditions for the fields, which are set above the minimum of the effective potential and subsequently oscillate around this minimum. The average behaviour of the scalar fields' energy densities is similar to the transient matter domination model. Initially there is a nearly tracking period at early times, followed by transition to nearly constant energy densities for the fields. Unlike the transient matter domination model, one of the CDM fluids then scales with the scalar fields' energy densities as shown in the right panel of Figure~\ref{backmatdom}. Although there is oscillatory behaviour at early times in the growth factor it does not exceed unity, and the average behaviour is very similar to that of the weaker coupled transient matter dominated model. As such the model  is consistent with present observations.

\begin{figure}
\centerline{\includegraphics[angle=0,height=80mm]{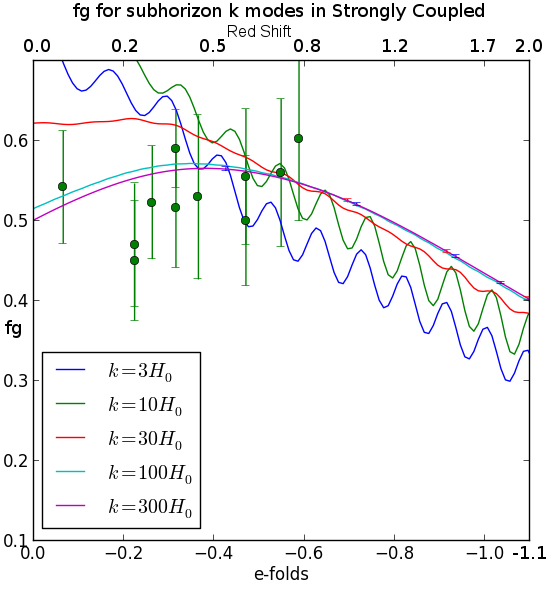}\includegraphics[angle=0,height=80mm]{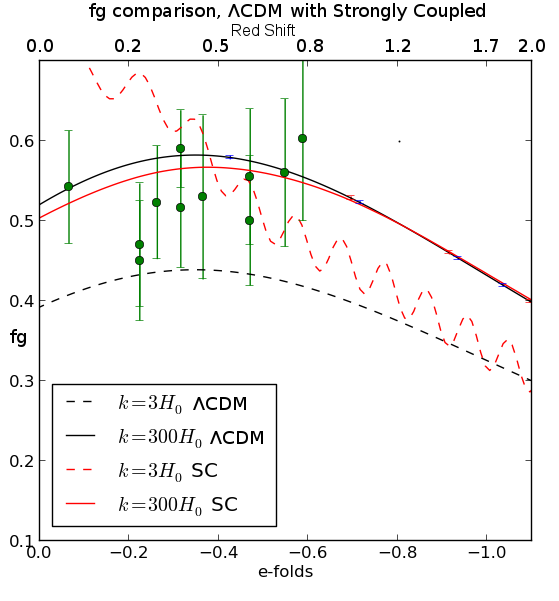}} 
\caption{The left plot shows the growth function, $fg$, sub-horizon scales, for strongly coupled matter dominated M$\ph$cQ model, for the region of redshifts relevant for current and future surveys. Couplings $\C_{11}=-0.2$, $\C_{12}=0.4$, $\C_{21}=-0.3$ and $\C_{22}=0.6$. Slopes for the potentials, $\lambda_1 = \lambda_2 = 0.1$. The right hand plot compares the $fg$ between $\Lambda$CDM and the strongly coupled model (SC) for $k=300H_0$ and $k=3H_0$.}
\label{NOfsubhorlong}
\end{figure}
\subsubsection{Scaling Solution}
\label{CQScaling}

\begin{figure}
\centerline{\includegraphics[angle=0,height=75mm]{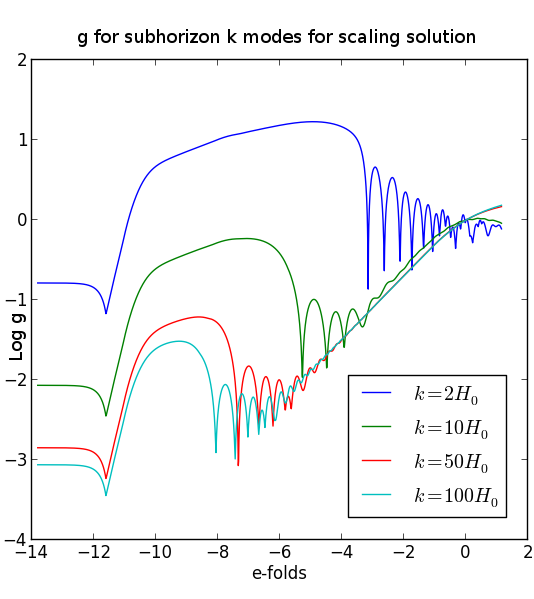}} 
\caption{The plot shows the log of growth factor, g, for scaling solution M$\ph$cQ model, for subhorizon $k$ modes.}
\label{Scalingg}
\end{figure}

As a second example we followed Ref.~\cite{Amendola:2014kwa}, and considered the same setup and potential, but chose couplings which give rise to a scaling behaviour. The resultant system is, however,  not consistent with observations. It even lacks dark matter domination at earlier epochs. In this example $\C_{11}=90$, $\C_{12}=-8$, $\C_{21}=-63$ and $\C_{22}=-10$ and  the slopes of the potentials were taken to be $\lambda_1 = 10 , \lambda_2 = 5.4$. For this example we calculated the growth factor, $g$, shown in Figure \ref{Scalingg}. It can clearly be seen that it becomes greater than unity on subhorizon scales, although less pronounced with increasing $k$, showing this model to be unrealistic at both the background and perturbed level.

\subsubsection{Exploration of Potential Slope Space for Strongly Coupled Matter Domination}
\label{Lspaceexplore}
\begin{figure}
\centerline{\includegraphics[angle=0,height=100mm]{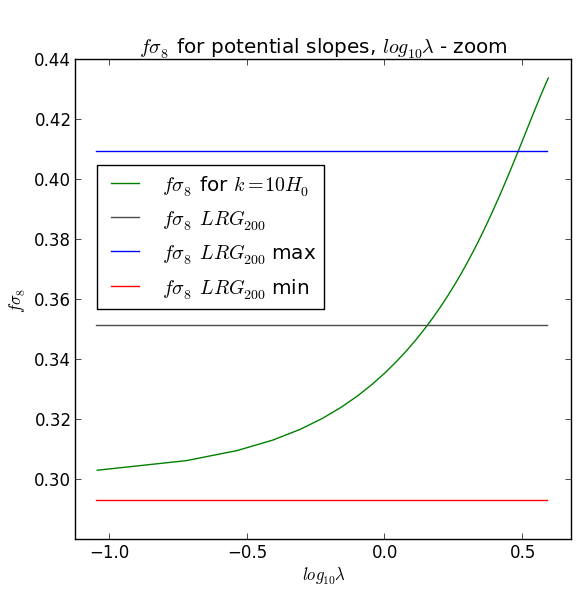}} 
\caption{$f\sigma_8$ for the matter dominated model with varying slopes for the potentials, $\lambda$. The wavenumber was set to $k=42H_0$ for these runs. Couplings, $\C_{11}=-20$, $\C_{12}=40$, $\C_{21}=-30$, $\C_{22}=60$. The observational values with uncertainties used for comparison were those from LRG$_{200}$, for $z=0.25$. The plot is a subsection from a region of $\lambda$ parameter space from $\lambda=10$ down to $\lambda=0.01$ where the results are consistent with observations.}
\label{NOslopesfs8COARSE}
\end{figure}

We now explore how changes in the slopes of the potentials (the $\lambda_I$ terms in \eq{sumofexppot}) in the matter dominated model affects the cosmology. Since, for the couplings in the strongly coupled model, the original large value of the slopes, $\lambda_1 = 10$, $\lambda_2 = 10$ produced excessive growth, we investigated the slope parameter space. This was done from $\lambda_I=10$ down to $\lambda_I=0.01$. This region including observationally consistent models is shown in Figure~\ref{NOslopesfs8COARSE}.

In producing this figure, the wavenumber of $k=42H_0$ was selected since it is the smallest $k$ mode for which SKA is predicted to still attain its highest precision~\cite{Bull:2015lja}. The LRG$_{200}$ data set was selected simply to serve as an example for comparison (see Section~\ref{Obs} for more details on  observations used for comparison). Different data sets would move the value of $f\sigma_8$ slightly, and alter the range of the error bars. There is a range of slopes for which these models not only gave a realistic background cosmology but also gave growth consistent with observations. In this region the parameter values are at least an order of magnitude smaller than the original values used. The background cosmologies for this region are very close in behaviour to Figure~\ref{backmatdom}. For slopes much smaller than $\lambda=0.1$ the potential is becoming increasingly flat and the results become noise dominated. As such they were excluded from our analysis.

\subsubsection{Exploration of Couplings Space for Strongly Coupled Matter Domination}
\label{Cspaceexplore}

For completeness a coarse exploration of the full parameter space of couplings was conducted and the growth function calculated. The range of couplings investigated was from $-50 \leq \C \leq 50$ with a stepping of $10$. The slopes for the potentials and initial conditions were left as before i.e. $\lambda_1 = \lambda_2 = 10$. For the portions of coupling space where the couplings satisfied the constraints for these models all exhibited excessive growth.\\

\begin{figure}
\centerline{\includegraphics[angle=0,height=100mm]{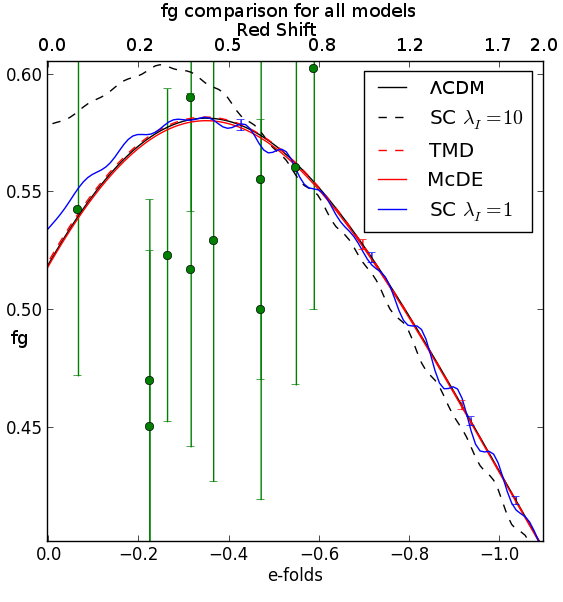}\includegraphics[angle=0,height=100mm]{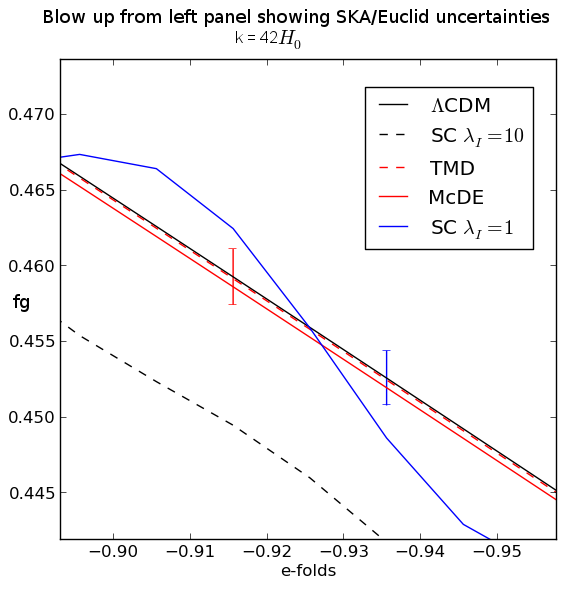}} 
\caption{This plot compares the $fg$ between $\Lambda$CDM and the strongly coupled model (SC) for both $\lambda_I=10$ and $\lambda_I=1$, the transient matter dominated model (TMD) and McDE model. All models are evaluated at $k=42H_0$. The insert zooms in on an example region in redshift space where future surveys should measure $fg$ sufficiently accurately to compare different model predictions.}
\label{allfg}
\end{figure}

Finally in Figure~\ref{allfg} we show $fg$ for a sample of the models studied against $\Lambda$CDM compared with the SKA and Euclid predicted precisions. This was carried out for mode $k=42H_0$ as it corresponds to the largest scale for which the highest predicted precision should be achieved for SKA~\cite{Bull:2015lja}. We can see that unless the best possible predicted precision is achieved it may still be hard to distinguish models with small couplings and slopes from $\Lambda$CDM. However, models with larger couplings should be easily identified. The strongly coupled model with $\lambda_I = 1$ was chosen since it lay within one of the viable regions discovered in Subsection~\ref{Lspaceexplore}. For this model it is clear that this would be distinguishable from $\Lambda$CDM given even the conservative predicted precision for SKA and Euclid. Therefore, there is a region of parameter space between the transient matter domination parameters and the strongly coupled parameters we initially tested in which subregions satisfy both background constraints and give growth results distinguishable from $\Lambda$CDM by future surveys, as the strongly coupled model does.

\section{Consistency test}
\label{GrowthvsObs}

The above analysis was performed assuming that the
current density parameters are the same as those obtained with Planck data
for a $\Lambda$CDM cosmology. This approach is simple to
implement but may lead to erroneous conclusions because a CMB fit to
interacting dark energy models can lead to different density
parameters of the various components than the ones obtained assuming
$\Lambda$CDM (see e.g.~Ref.~\cite{Valiviita:2015dfa}). See also
Appendix A of~Ref.~\cite{Jennings:2009qg} for a succinct but detailed
discussion on this topic.

To ensure the validity of our results, therefore, we undertake a
consistency test to confirm that the models studied are sufficiently
close to $\Lambda$CDM at the time of decoupling and consequently that
their background evolutions yield negligible differences. This is of
course not entirely sufficient as there is also some contribution from the late
integrated Sachs-Wolfe effect brought in by the very recent dark
energy domination, nonetheless, the exercise should be
informative, given that a full parameter constraint analysis is beyond
the scope of this work.

In practice we compared for our different models the value of $\Omega_M$ at
the time of decoupling with the same quantity for $\Lambda$CDM, and
evaluated whether there is any significant deviation. In the cases
where there is a discrepancy we do expect a change in the value of the
current energy densities had a CMB fit been carried out. Moreover, a
change in the values of the growth factor and of $fg$ is also to be
expected. 
In this case it is not possible to trust the naive comparison of the results of our study 
with 
survey data processed 
assuming $\Lambda$CDM, and a full parameter fit of background and perturbed parameters  
would be required.

\begin{figure}
\centerline{\includegraphics[angle=0,height=100mm]{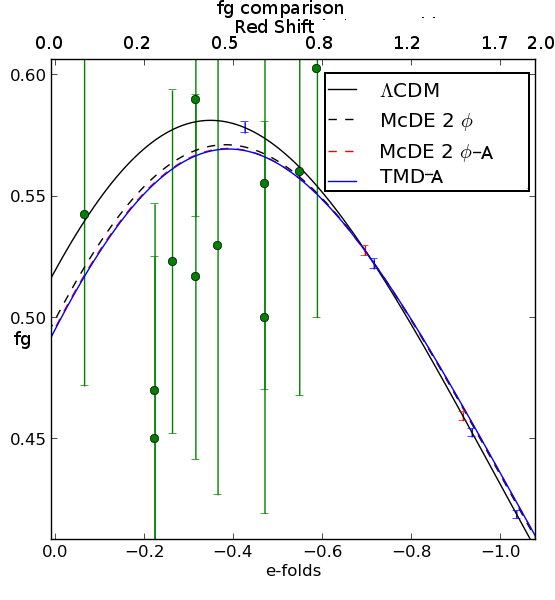}\includegraphics[angle=0,height=100mm]{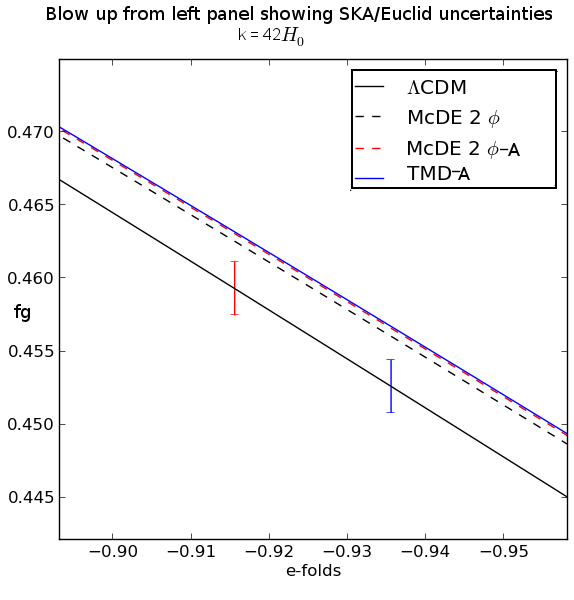}} 
\caption{These plots compare the $fg$ between $\Lambda$CDM and various iterations of the TMD and McDE models. All models are evaluated at $k=42H_0$. The right hand plot zooms in on an example region in redshift space where future surveys should measure $fg$ sufficiently accurately to compare different model predictions. The black solid line is for McDE as defined in Section~\ref{McDE} but with an additional field with the same size slopes and couplings to the matter species. The red dashed line is the same model with larger couplings of $\pm 0.095$ and slopes of $\lambda_I = 0.1$. Finally the blue solid line is a ``balanced" TMD model with couplings of +0.07, -0.07, +0.12 and -0.12 and $\lambda_I = 0.1$.}
\label{fgGrowthvsObs}
\end{figure}

Given that the McDE model of Section~\ref{McDE} has $\Omega_M$ close
to that of $\Lambda$CDM at decoupling ($<1\%$ deviation), we used this
set up as a starting point to investigate the effects of multiple
fields on the growth of structure. We show the results in
Figure~\ref{fgGrowthvsObs}.  The black dashed line shows an extension
to the McDE model, named ``McDE 2 $\phi$'', where an additional field was
added, but with the same magnitude for the couplings and slopes as in the
original McDE model in Section~\ref{McDE}, i.e.~$C_{11} =
0.024$, $C_{12}=-0.024$. By comparing with Figure~\ref{allfg},
we see that this modification already gives a larger difference in the
growth from $\Lambda$CDM than for the original McDE model.  However,
although the value of $\Omega_M$ at decoupling deviates from the one
of the $\Lambda$CDM model by less than $1\%$, it is not as close to
$\Lambda$CDM as the original McDE model, and for the purposes of the
validity of our analysis we seek models which match $\Lambda$CDM
$\Omega_M$ at decoupling exactly. This was achieved by adjusting the
couplings of the McDE 2 $\phi$ model such that $C_{11}=C_{21} =
0.095$, $C_{12}=C_{22}=-0.095$ and $\lambda_I = 0.1$. This model,
named ``McDE 2 $\phi$-A'', corresponds to the red dashed line in
Figure~\ref{fgGrowthvsObs}.  While this has the same value for $\Omega_M$
at decoupling to $\Lambda$CDM, the growth deviated from this by more
than the SKA and Euclid uncertainties at the redshifts shown.
 
As a third example we have considered a modified TMD model. By this we
mean that for each CDM species coupling to a given field the couplings
are of equal magnitude and opposite sign, but differ in magnitude
between the fields. We named this model ``TMD-A'', represented with a blue
solid line in Figure~\ref{fgGrowthvsObs}. More specifically, the
couplings are $C_{11}=0.07$, $C_{12}= -0.07$, $C_{21}=0.12$ and
$C_{22}=-0.12$ and $\lambda_I = 0.1$. This also gives identical
$\Omega_M$ at decoupling to $\Lambda$CDM, while the growth deviates
from the growth in $\Lambda$CDM by more than the SKA and Euclid
uncertainties at the redshifts shown, and slightly more than all the
other models shown. The SKA and Euclid uncertainties are themselves at
the $1\%$ level and we are therefore confident that the growth for
these models should be distinguishable from that predicted by
$\Lambda$CDM by these future surveys.

\section{Discussion and conclusion}
\label{Conclusion}

In this paper we have presented the full equations for perturbations in M$\ph$cQ models, produced a numerical package to evolve these perturbations, \PY, and used this package to compare a set of example models with observations. We found that the longitudinal gauge, often employed in previous studies of less general systems, is not ideal for the numerical evolution of the full system, and we therefore used the flat gauge.

We found that there are examples of M$\ph$cQ models which lie within current observational bounds, however,  distinguishable from $\Lambda$CDM models with future surveys such as Euclid and SKA, as they will attain a precision in $fg$ at the percent level or better \cite{Raccanelli:2015qqa}. On the other hand, we also found examples such as the strongly coupled model defined in Ref.~\cite{Amendola:2014kwa}, were $fg$ is incompatible with current observations, ruling out the model. This confirms  the conclusion in Ref.~\cite{Amendola:2014kwa}, that while ``large'' couplings might give a realistic background model, the perturbations experience excessively strong growth (or damping) and are, therefore, unrealistic. However, we found that it did not require both the couplings and the slopes to be reduced simultaneously in order for a region of viable background and perturbed cosmologies to be recovered, as discussed in Subsection~\ref{Lspaceexplore}, since when $\lambda \lesssim 2$ this leads to a viable parameter space region.

We have found for the McDE model, and the transient matter dominated case for the M$\ph$cQ models studied, that they give realistic background cosmologies while apparently exceeding the allowed coupling strength for single field M$\ph$cQ, $\C \lesssim 0.1 \sqrt{\frac{2}{3}}$ (see e.g.~Ref.~\cite{Amendola:1999er}). This difference in behaviour between single field (and single CDM species) and multiple CDM species models results from the relative signs of the couplings. In Ref.~\cite{Piloyan:2014gta}, the McDE model with couplings significantly greater than $0.1 \sqrt{\frac{2}{3}}$ gave rise to viable background and perturbed cosmologies. This is attributed to the unique way in which the CDM species are oppositely charged with respect to the DE scalar field (couplings are also of the same magnitude). In our M$\ph$cQ models each CDM species has an opposite charge relative to each scalar field i.e.~CDM species 1 has a negative coupling to scalar field 1 while CDM species 2 has a positive coupling, and similarly for scalar field 2. Although the couplings are no longer symmetric in magnitude, this partial balance of charge still has a similar effect as in McDE, both in giving viable background cosmologies and in controlling the growth of structure. However, of the models studied only the transient matter dominated model satisfied both the background evolution and the evolution of growth through $fg$ for low redshift.

Finally, we have also addressed the question of the applicability of the large $k$ approximation, and investigated at which scales it may be considered a good approximation. The deviation of the full equation results for large $k$ modes from the approximation is frequently greater than the experimental uncertainty in future surveys. In Section~\ref{McDE} we showed that using a subhorizon approximation gave a difference in results for growth from the full equations which would be larger than the predicted observational precision for SKA and Euclid. The approximation already deviates from the full equations by more than the predicted precision of SKA~\cite{Bull:2015lja} at $k=300H_0$ and becomes progressively worse towards $k=42H_0$, the boundary for which SKA is predicted to have the highest precision. Hence results from the full equations should be used for comparison with future observations instead of those obtained using the approximation. This is therefore an important aspect to take into account in the analysis of large scale structure from near future experiments.

\begin{acknowledgments}
The authors are grateful to Ian Huston, Pedro Carrilho, Phil Bull and Shailee Imrith for helpful discussions. AL is funded by a
STFC studentship, KAM is supported, in part, by STFC grant
ST/M001202/1 and DJM is supported by a Royal Society University Research Fellowship. N.J.N thanks Queen Mary University of London for hospitality. His research work is supported by the grant UID/FIS/04434/2013.
The computer algebra package
{\sc{Cadabra}}\cite{Cadabra} was used in the derivation of some of the
equations.
\end{acknowledgments}

\appendix

\section{Gauge Transformations}
\label{Gauge Transformations}

\subsection{General Gauge Transformations}
\label{General Gauge Transformations}

We now give the gauge transformations for the perturbed quantities
used in the main body of this paper and in Subsection~\ref{Flat to
  Long Gauge Relations} below. Following the notation of
Ref.~\cite{Malik:2004tf}, quantities in the new coordinate system are
  denoted by a tilde.

The matter variables, the velocity and the density perturbations,
transform as
\bea
\label{GGTV}
\tilde{\hat v}_\alpha &=& \hat v_\alpha + \frac{\delta t}{a} \,,\\
\label{GGTrho}
\tilde{\delta \rho}_\alpha 
&=& \delta \rho_\alpha - \dot{\bar{\rho}}_\alpha \delta t \,,
\eea
where $\hat v_\alpha$ is defined in \eq{FlatNewv}.

The perturbations of the metric transform as
\bea
\label{GGTphi}
\tilde{\phi} &=& \phi - \dot{\delta t} \,,\\
\label{GGTpsi}
\tilde{\psi} &=& \psi + H \delta t \,,\\
\tilde{B} &=& B - a\dot{\delta x}+ \delta t \,,\\
\tilde{E} &=& E - \delta x \,.
\eea

\subsection{Flat to Longitudinal Gauge Relations}
\label{Flat to Long Gauge Relations}

The relation between the velocity in flat and in longitudinal gauge is given by
\be
\label{FLGTV}
\hat{v}_{\alpha(\rm flat)} = v_{\alpha(\rm long)} + {B}_{(\rm flat)}\,.
\ee
The relation for the  density perturbations is
\be
\label{FLGTrho}
{\delta \rho}_{\alpha(\rm flat)} = \delta \rho_{\alpha(\rm long)} - a \dot{\bar{\rho}}_\alpha {B}_{(\rm flat)} \,.
\ee
The transformation behaviour of the metric perturbations and the fact that
$\phi = \psi$ in longitudinal gauge in the absence of anisotropic
stress gives 
\be
\label{FLGTpsi}
{B}_{(\rm flat)} = - \frac{\phi_{(\rm long)}}{Ha} \,.
\ee

\section{Longitudinal Gauge with with Arbitrary Numbers of Fields and DM Fluids}
\label{Long2f2dm}

As mentioned in Section~\ref{Fixing} the \PY~code was originally written in longitudinal gauge as this is one commonly used in literature in the field, see e.g.~\cite{Amendola:2014kwa}. However due to numerical instabilities caused by the constraint \eq{Phiconstraint} for $\phi$ below, this version was abandoned. We include the equations below for reference and completeness.

For a given DM species, $\alpha$, the evolution equation for the perturbation is
\be
\label{IntDEPertEcons4}
\dot{\delta \rho_\alpha} + 3 H (\delta \rho_\alpha + \delta P_\alpha) - \left(3 \dot{\phi} + \frac{k^2 v_\alpha}{a}\right)({\bar{\rho}}_\alpha + {\bar{P}}_\alpha) = - \sum\limits_{I} \kappa \C_{I \alpha} ({\bar{\rho}}_\alpha - 3 {\bar{P}}_\alpha) {\dot{\delta \ph}}_I - \sum\limits_{I} \kappa \C_{I \alpha} (\delta \rho_\alpha - 3 \delta P_\alpha) {\dot{\bar{\ph}}}_I .
\ee
Momentum conservation is given by

\be
\label{IntDEPertMomconsLong}
\dot{v}_\alpha =  \kappa \sum\limits_{I} \C_{I \alpha} (\bar{\rho}_\alpha - 3 \bar{P}_\alpha) \frac{\delta \ph_I}{a} + 3H \frac{\dot{\bar{P}}_\alpha}{\dot{\bar{\rho}}_\alpha} (v_\alpha) - H(v_\alpha) - \frac{\phi}{a} - \frac{\delta P_\alpha}{a({\bar{\rho}_\alpha} + \bar{P}_\alpha)} .
\ee

The evolution equation for the fields, labelled $I$, $J$, is
\be
\label{IntDEPertEconsSF2}
{\ddot{\delta \ph}}_I + 3 H {\dot{\delta \ph}}_I + \sum\limits_{J} V,_{\ph_I \ph_J} \delta \ph_J - 4 \dot{\phi} {\dot{\bar{\ph}}}_I + \frac{k^2}{a^2} \delta \ph_I + 2 V,_{\ph_I} \phi - 2 \sum\limits_{\alpha} \kappa \C_{I \alpha} ({\bar{\rho}}_\alpha - 3 {\bar{P}}_\alpha) \phi - \sum\limits_{\alpha} \kappa \C_{I \alpha} (\delta \rho_\alpha - 3 \delta P_\alpha) = 0 . 
\ee
The Einstein Field Equations are as follows. From the $0-0$ component we get
\be
\label{G00DEEFE2}
3 H (\dot{\phi} + H \phi) + \frac{k^2}{a^2}\phi = - \frac{\kappa^2}{2} \left[\sum\limits_{\alpha} \delta \rho_\alpha + \sum\limits_{I}(-\phi \dot{\bar{\ph}}^2_I + {\dot{\delta \ph}}_I {\dot{\bar{\ph}}}_I + V,_{\ph_I} \delta \ph_I)\right] .
\ee
From the $0-i$ component we get
\be
\label{Gi0DEEFE2}
\dot{\phi} + H \phi = - \frac{\kappa^2}{2} \left[\sum\limits_{\alpha} a v_\alpha(\bar{\rho}_\alpha + \bar{P}_\alpha) - \sum\limits_{I} {\dot{\bar{\ph}}}_I \delta \ph_I \right] .
\ee
From the trace of $i-j$ component we get
\be
\label{TraceGijDEEFE2}
\ddot{\phi} + 4 H \dot{\phi} + (3 H^2 + 2 \dot{H}) \phi = \frac{\kappa^2}{2} \left[\sum\limits_{\alpha} \delta P_\alpha - \sum\limits_{I} \left(\phi \dot{\bar{\ph}}^2_I - {\dot{\delta \ph}}_I {\dot{\bar{\ph}}}_I + V,_{\ph_I} \delta \ph_I\right) \right] .
\ee
From the trace-free part of the $i-j$ component we get
\be
\label{TraceFreeGijDEEFE2}
\psi = \phi ,
\ee
since $\sigma_s = 0$.\\
From \eq{G00DEEFE2} and \eq{Gi0DEEFE2} we get

\be
\label{Phiconstraint}
\phi = \left( \sum\limits_{I} \dot{\bar{\ph}}^2_I - \frac{2 k^2}{(\kappa a)^2} \right)^{-1} \left[ \sum\limits_{\alpha} \left( \delta \rho_\alpha - 3 H a v_\alpha(\bar{\rho}_\alpha + \bar{P}_\alpha) \right) + \sum\limits_{I} \left( {\delta \dot{\ph}_I \dot{\bar{\ph}}_I} + {V,_{\ph_I} \delta \ph_I} + 3 H \dot{\bar{\ph}}_I \delta \ph_I \right) \right]
\ee

\section{Synchronous Comoving Gauge with Arbitrary Numbers of Fields and DM Fluids}
\label{Synch2f2dm}
Synchronous gauge had been considered for use in the \PY~code. This was partly because it has been used in codes such as CAMB and CLASS~\cite{Lewis:1999bs,Blas:2011rf}. The equations from Section~\ref{IntDEperts} are presented here in synchronous co-moving gauge ($\tilde{\phi} = \tilde{B} = \tilde{v} = 0$), but otherwise in full generality, allowing for multiple fields and fluids. This is done for reference and completeness. For a given DM species, $\alpha$, the evolution equation for the perturbation is

\be
\label{IntDEPertEconsSynch}
\dot{\delta \rho_\alpha} + 3 H (\delta \rho_\alpha + \delta P_\alpha) - \left(3 \dot{\psi} + k^2 \dot{E}\right)({\bar{\rho}}_\alpha + {\bar{P}}_\alpha) = - \sum\limits_{I} \kappa \C_{I \alpha} ({\bar{\rho}}_\alpha - 3 {\bar{P}}_\alpha) {\dot{\delta \ph}}_I - \sum\limits_{I} \kappa \C_{I \alpha} (\delta \rho_\alpha - 3 \delta P_\alpha) {\dot{\bar{\ph}}}_I .
\ee
Momentum conservation is given by
\be
\label{IntDEPertMomconsSynch}
 \kappa \sum\limits_{I} \C_{I \alpha} ({\bar{\rho}}_\alpha - 3 \bar{P_\alpha}) \delta \ph_I  = \frac{\delta P_\alpha}{{\bar{\rho}}_\alpha + \bar{P_\alpha}} .
\ee
The evolution equation for the fields, labelled $I$, $J$, is
\be
\label{IntDEPertEconsSFSynch}
{\ddot{\delta \ph}}_I + 3 H {\dot{\delta \ph}}_I + \sum\limits_{J} V,_{\ph_I \ph_J} \delta \ph_J - \left(3 \dot{\psi} + k^2 \dot{E}\right) {\dot{\bar{\ph}}}_I + \frac{k^2}{a^2} \delta \ph_I - \sum\limits_{\alpha} \kappa \C_{I \alpha} (\delta \rho_\alpha - 3 \delta P_\alpha) - 2 \kappa \sum\limits_{\alpha} \C_{I \alpha} \bar{\rho}_{\alpha} = 0 . 
\ee
The Einstein Field Equations are as follows. From the $0-0$ component we get
\be
\label{G00DEEFESynch}
3 H (\dot{\psi})+ \frac{k^2}{a^2} (\psi + H a^2 \dot{E} ) = - \frac{\kappa^2}{2} \left[\sum\limits_{\alpha} \delta \rho_\alpha + \sum\limits_{I}({\dot{\delta \ph}}_I {\dot{\bar{\ph}}}_I + V,_{\ph_I} \delta \ph_I)\right] .
\ee
From the $0-i$ component we get
\be
\label{Gi0DEEFESynch}
\dot{\psi} =  \frac{\kappa^2}{2} \sum\limits_{I} {\dot{\bar{\ph}}}_I \delta \ph_I .
\ee
From the trace of $i-j$  component we get
\be
\label{TraceGijDEEFESynch}
\ddot{\psi} + 3 H \dot{\psi} = \frac{\kappa^2}{2} \left[\sum\limits_{\alpha} \delta P_\alpha + \sum\limits_{I} \left({\dot{\delta \ph}}_I {\dot{\bar{\ph}}}_I - V,_{\ph_I} \delta \ph_I\right) \right] .
\ee
From the trace-free part of the $i-j$  component we get
\be
\label{TraceFreeGijDEEFESynch}
{\dot{\sigma}}_s + H \sigma_s + \psi = 0 ,
\ee
where $\sigma_s$ is the scalar shear and $\sigma_s = a^2 \dot{E}$.




\end{document}